# Understanding the uneven spread of COVID-19 in the context of the global interconnected economy

**Dimitrios Tsiotas[1,2*] and Vassilis Tselios[3]**
1. Department of Regional and Economic Development, Agricultural University of Athens, Greece, Nea Poli, Amfissa, 33100, Greece.
2. Adjunct Academic Staff, Hellenic Open University, Athens, 10677, Greece
3. Department of Economic and Regional Development, Panteion University of Social and Political Sciences, Athens, 17671, Greece.
Emails: tsiotas@aua.gr – tsiotas.dimitrios@ac.eap.gr; v.tselios@panteion.gr
* Corresponding author

**Abstract**
Using network analysis, this paper develops a multidimensional methodological framework for understanding the uneven (cross-country) spread of COVID-19 in the context of the global interconnected economy. The globally interconnected system of tourism mobility is modeled as a complex network, where two main stages in the temporal spread of COVID-19 are revealed and defined by the cutting-point of the 44[th] day from Wuhan. The first stage describes the outbreak in Asia and North America, the second one in Europe, South America, and Africa, while the outbreak in Oceania is spread along both stages. The analysis shows that highly connected nodes in the global tourism network (GTN) are infected early by the pandemic, while nodes of lower connectivity are late infected. Moreover, countries with the same network centrality as China were early infected on average by COVID-19. The paper also finds that network interconnectedness, economic openness, and transport integration are key determinants in the early global spread of the pandemic, and it reveals that the spatio-temporal patterns of the worldwide spread of COVID-19 are more a matter of network interconnectivity than of spatial proximity.

**1. INTRODUCTION**
The COVID-19 coronavirus disease (SARS-CoV-2) is the new pandemic that emerged in December 2019, in Wuhan city, China, and has ever since been rapidly spread around the world, causing cascading deaths to humanity, vast pressures on the national systems of public health, and uncertainty about the future of the global and national economies (Anderson et al., 2020; WHO, 2020; Wu et al., 2020). Although it has been only about one year since its emergence in Wuhan, the COVID-19 pandemic has already become a major concern and priority for the scientific community today (Harapan et al., 2020). According to the Google Scholar academic database (Google Scholar, 2020), the search of the keyword "COVID-19" already yields approximately 1.34m (million) results, whereas other keywords referring to established research fields (many of which enjoy centuries of scientific research) yield a comparable number of results, such as the keywords "gravity" (3.9m results), "cancer" (4.6m results), "networks" (5.33m results), electric (5.93m results), health (6.87m results), "space" (7.06m results) and "science" (8.38m results). The peculiar biological and epidemiologic features of COVID-19, such as its potential to be easily transmitted through respiratory channels (Xu et al., 2020), its ability to infect healthy people through both ill and asymptomatic patients (Fang et al., 2020; Heymann and Shindo, 2020), its considerably long incubation time (Briz-Redon and Serrano-Aroca, 2020; Xu et al., 2020), along with many other features concerning its propagation ability, biological resilience, and massiveness (Chen et al., 2020; Flaxman et al., 2020), have equipped this disease with threat-properties, causing the global emergence. In addition to these biological and epidemiologic features, provided that at the macroscopic (or macroeconomic) level, the defining characteristic of a pandemic is related to the ability to spread rapidly throughout the world (Flaxman et al., 2020; Rossman et al., 2020; Ruktanonchai et al., 2020; Oliveira et al., 2021), the already fatal worldwide outcome of COVID-19 (WHO, 2020) has allowed this disease to meet the requirements for it to be considered a pandemic (Demertzis et al., 2020).

In the already vast literature on COVID-19, at a glance it can be observed that there are three major research directions (strands) in the study of the pandemic's spread: the first concerns the patterns and causes of the spread (e.g. Arpino et al., 2020; Menkir et al., 2020; Vespignani et al., 2020; Wang et al., 2020), the second one its effects (e.g. Bonaccorsi et al., 2020; Betsch et al., 2020; Chinazzi et al., 2020; Flaxman et al., 2020), and the third one its management, treatment and cure (e.g. Karatayev et al., 2020; Spelta et al., 2020; Weitz et al., 2020). These directions can involve various applications, such as in the fields of individual and public health, medicine and clinical research, pharmacy, policy, economy, society, education, communication, transportation, environment, geography, and many others. As is evident from all these diverse approaches, scientific research on COVID-19 consists of a complex and multidisciplinary context (Chen et al., 2020; Vespignani et al., 2020; Xu et al., 2020), which denotes the desire of the whole scientific community to deal with this issue in all aspects of everyday life, aiming to get ahead in the fight against the pandemic. Although a thorough review of the vast and multidisciplinary COVID-19 literature is a challenge for future researchers of epistemology, a prime observation that is apparent to every scholar and researcher dealing with this issue regards the symbiotic relationship between the spread of the pandemic and the interconnectedness of the modern world that supports the spread process.

In epidemiological terms, the infection of COVID-19 is possible through the transmission of the virus between two individuals (Demertzis et al., 2020; Fang et al., 2020; Heymann and Shindo, 2020; Xu et al., 2020; WHO, 2020), which in terms of communication theory (McQuail, 1987; Griffin, 2006) is interpreted either as an act of signal transmission, or flow-transferring or communication in general. Given that a pair-wise signal-transmission describes an elementary structure of connectivity, which in terms of the network paradigm is called link or edge (Christakis and Fowler, 2009; Newman, 2010; Barabasi, 2013), the virus transmission of COVID-19 can excellently be comprehended within the context of connectivity, as is outlined by network science (Newman, 2010; Barabasi, 2013; Barthelemy, 2011; Brandes et al., 2013). Therefore, the interconnectedness, which describes all aspects of everyday life, can be interpreted as the prime and major determinant of the spread of any pandemic, at any spatial and functional level. At the microscopic level, the interrelation between interconnectedness across individuals and the spread (transmission) of COVID-19 is mainly studied from the clinical and epidemiologic perspective, and it has already enjoyed fruitful research (Fang et al., 2020; Heymann and Shindo, 2020; Lescure et al., 2020), which allowed the scientific community to understand the pandemic, and thus to move forward towards the development of vaccinations and other treatments (Ahmed et al., 2020; Jeyanathan et al., 2020; Saad-Roy et al., 2020).

In macroscopic terms, this bipolar spread-interconnectedness relation can be comprehended on a dual basis. The first interprets its components within the context of countries (regardless of the geographical scale), whereas the second one conceptualizes them in a cross-country (between countries) framework. The within-countries consideration already involves fruitful research covering all topics of interest about the pandemic, such as regional outbreaks and spread due to imported cases (Menkir et al., 2020; Rossman et al., 2020), mobility and travel restrictions (Bonaccorsi et al., 2020; Chinazzi et al., 2020), the effects of lockdown (Gatto et al., 2020; Karatayev et al., 2020), and others (Arpino et al., 2020; Ruktanonchai et al., 2020). Even when it is implemented on an international scale, this approach conceptualizes interconnectedness as an intrinsic property of countries and interprets the uneven spread of the pandemic on the basis of comparing such intrinsic properties between countries. However, interconnectedness is broader than a within-country concept because it also concerns connectivity, which is developed in a cross-country context. The cross-country approach can be considered as more relevant in describing the real interconnected structure of the real-world because it conceptualizes the spread-channels (links) of the pandemic, not only within but also between countries, and thus it can provide insights towards a better understanding of the uneven spread of COVID-19 across countries. Despite its importance, this approach has not yet been examined within a comprehensive context, because current studies dealing with this issue (Arpino et al., 2020; Farzanegan et al., 2020; Hafner, 2020; Kapitsinis, 2020) do not conceptualize global interconnectedness within the context of network science (Newman, 2010; Barabasi, 2013; Barthelemy, 2011), which provides integrated modeling of communication systems.

Aiming to respond to this demand, this paper focuses on the patterns and causes (first strand) of the spread of COVID-19 and develops a multidimensional methodological framework for understanding the spatio-temporal spread of the pandemic in the context of the global economy, which is modeled as an interconnected cross-country structure. To do so, it conceptualizes global interconnectedness based on economic globalization (Shrestha et al., 2020) and, specifically, by constructing a network model of international tourism flows, which are considered as a good proxy (Farzanegan et al., 2020) for the outbreak of the pandemic. The study builds on a three-dimensional conceptual model for the analysis of the worldwide spatio-temporal spread of COVID-19 and incorporates one dimension approximating the interconnectedness of the international tourist mobility, a second one describing the openness of countries to the globalized economy, and a third one expressing the spatial impedance to transportation. By constructing a single network model, this paper proposes an integrated framework for the study of the spatio-temporal spread of COVID-19 and to contribute to the literature with more realistic models of the worldwide interconnected system, where COVID-19 and other pandemics are spread.

The remainder of this paper is organized as follows: Section 2 presents the methodological and conceptual framework of the study, Section 3 shows the results of the analysis and discusses them within the context of regional and geographical sciences, and finally, in section 4 conclusions are given.

## 2. METHODOLOGY AND DATA

This paper develops a multidimensional methodological framework for understanding the uneven (cross-country) spread of COVID-19 in the context of the global interconnected economy. The temporal spread of the pandemic is measured in terms of the time at which the first infection emerged from Wuhan to each country, which is considered as a good proxy for interconnectedness because it quantifies the impedance of the pandemic flows between countries (i.e. the level at which each international link resists or facilitates the transmission of the pandemic flows in the network). The methodological framework of the study builds on a three-dimensional (3D) conceptual model for the analysis of the worldwide spatio-temporal spread of COVID-19. The 3D-model incorporates one dimension (component) approximating the interconnectedness of international human mobility (1D: global network interconnectedness), a second one describing the economic structure and openness of countries to the globalized economy (2D: economic structure and openness), and a third one expressing the spatial impedance to transportation (3D: spatial impedance). This 3D conceptual model is developed to detect whether patterns in the cross-country temporal spread of COVID-19 can be related to various aspects of interconnectedness in the structure of the worldwide mobility system and, thus, to provide insights into the topological, socioeconomic, and geographical factors which affect the uneven spread of the pandemic.

Within this context, to study the international spread of COVID-19, we construct and collect twenty-four (24) variables (Table 1), the first of which includes epidemiologic information referring to the time-distance (measured in days from Wuhan - dfW) of the first confirmed case (infection) per country, whereas the other 23 variables are grouped into the categories of the overall 3D conceptual approach. For the configuration of the variables included in the 1st category (global network interconnectedness), the analysis employs graph modeling to represent the globally interconnected system of tourism mobility as a complex network. The graph-model is built on data referring to the year 2018, which were extracted from the Organization for Economic Co-operation and Development (OECD, 2021) and include records of the top-5 markets (including either OECD or non-OECD countries) of the inbound and outbound tourism flows per OECD country. Based on the available data, a directed weighted graph $G(V,E)$ of the global network of tourism flows (Global Tourism Network - GTN) was constructed, where nodes ($i$) correspond to tourism-destination countries and links ($ij$) to the annual number of tourists originating from a node (country of origin) $i \in V$ and visited node $j \in V$ (destination country). The GTN is a connected graph consisting of $n=75$ nodes and $m=179$ links (edges) that is modeled in the *L*-space representation (see Barthelemy, 2011), where nodes are geo-referenced at the coordinates of the countries' capital cities by using the Web Mercator projection (Google Maps, 2020).

**Table 1**

Variables participating in the analysis of COVID-19 global spatio-temporal spread

| Group | Symbol | Description | Source/Reference |
|---|---|---|---|
| EPIDEMICS | DFW | **Days from Wuhan:** The number of days from Wuhan since the first emergence of COVID-19 infection-case, per country (measured in dfW). | Worldometers (2020) |
| 1D GLOBAL NETWORK INTERCONNECTEDNESS | DEG | **Node degree:** The number of edges that are adjacent to a given node of a network. This expresses the node's communication potential. | Newman (2020)[*] |
| | IN.DEG | **Node in-degree:** The number of incoming connections that are adjacent to a given node. | |
| | OUT.DEG | **Node out-degree:** The number of outgoing connections that are adjacent to a given node. | |
| | STR | **Node strength:** Defined by the sum of edge-weights that are adjacent to a given network node. | |
| | IN.STR | **Node in-strength:** The sum of weights of the incoming connections that are adjacent to a given node. | |
| | OUT.STR | **Node out-strength:** The sum of weights of the outgoing connections that are adjacent to a given node. | |
| | C | **Node clustering coefficient:** The probability of meeting linked neighbors around a node, which is defined by the number of the node's connected neighbors divided by the number of the total triplets shaped by this node. | Koschutzki et al. (2005)[*] |

| Group | Symbol | Description | Source/Reference |
|---|---|---|---|
| | CB | **Node betweenness centrality:** The proportion defined by the number of the shortest-paths passing through a given node to the total number of the network shortest-paths. | |
| | CC | **Node closeness centrality:** The inverse of the total binary distance computed on the shortest paths originating from a given node having the other nodes in the network as destinations. This expresses the node's reachability from the other nodes in the network. | |
| | ECC | **Node eccentricity:** The maximum network-distance from a given node to all the others in the network. | |
| **2D ECONOMIC STRUCTURE AND OPENNESS** | GI | **Overall KOF Globalisation Index:** Composite index measuring economic, social, and political globalization, yearly, from 1970 to 2017. Data refer to the year 2017. | ETH Zurich KOF (2019) |
| | GDP | **Gross Domestic Product (GDP):** GDP is the sum of gross value added by all resident producers in the economy plus any product taxes and minus any subsidies not included in the value of the products. It is calculated for the year 2017, without making deductions for depreciation of fabricated assets or depletion and degradation of natural resources. Data are in constant 2010 U.S. dollars. | Worldbank (2020) |
| | TFP | **Total factor productivity (TFP):** Aggregate indicator (loosely) expressing the amount of growth achieved due to both labor and capital productivity factors. It is computed on constant national prices (2011=1). | |
| | POP | **Population:** The number of citizens of the country according to the most recent national census. | GGDC (2020) |
| | HC | **Human Capital:** Human capital index, based on years of schooling and returns on education. | |
| | GDP.pc | **GDP per capita:** The GDP divided by mid-year population. Data are in constant 2010 U.S. dollars. | Worldbank (2020) |
| | TFP.pc | **Total factor productivity per capita:** The TFP, which is divided by the country's population. | GGDC (2020) |
| **3D SPATIAL IMPEDANCE** | CST | **Coastal indicator:** Dummy (binary indicator) variable indicating whether a country is coastal (1) or not (0). | Own elaboration, based on Google Maps (2020). |
| | DSTFC | **Distance from China:** The shortest geographical distance of a country from China (measured in km). | |
| | RDL | **Road length:** The length of the road network in each country (measured in km). | Citypopulation (2020) |
| | RLL | **Rail length:** The length of the rail network in each country (measured in km). | Nationmaster (2020) |
| | PRT | **Ports:** The number of active ports in each country, for the year 2020. | Worldportsource (2020) |
| | APRT | **Airports:** The number of active airports in each country, for the year 2020. | Globalfirepower (2020) |

*. Own elaboration for the GTN, based on OECD (2021) database for the year 2018.

The variables included in the other two categories (2D: economic openness, and 3D: spatial impedance) are extracted from various web-sources of secondary data (Citypopulation, 2020; ETH Zurich KOF, 2019; GGDC, 2020; Globalfirepower, 2020; Google Maps, 2020; Nationmaster, 2020; Worldbank, 2020; Worldometers, 2020; Worldportsource, 2020), where cases only referring to the countries included in the GTN are included in the variables' configuration. Within this context, all the available variables of Table 1 are of length 75, with each element referring to a GTN node (country). Based on these 24 variables, this paper builds on a multidimensional network analysis employing methods of statistical mechanics, such as descriptive and statistical-inference analysis, parametric fitting, and non-parametric estimation methods (Botev et al., 2010; Kim and Scott, 2012; Walpole et al., 2012; Tsiotas, 2019, 2020), to study the uneven spread of the COVID-19 pandemic. Network analysis (also called complex network analysis or network science, depending on whether it is seen as a discipline, see Barthelemy, 2011; Barabasi, 2013; Brandes et al., 2013) uses the network paradigm to represent complex communication (or other interacting) systems to graphs, which are pair-sets of nodes and edges that are modeled in terms of adjacency, connectivity and weights matrices (Newman, 2010). In comparison with other models of socioeconomic or spatial interaction, graphs have the advantage of including, in a single model, both structural and functional information (Easley and Kleinberg, 2010), which is available both on a local (i.e. node, neighborhood) and global (i.e. network) scale and therefore to provide a double (hybrid) microeconomic and macroeconomic description of the real-world systems (Tsiotas and Polyzos, 2018). Building on the network paradigm, this paper approaches the concept of global interconnectedness within this hybrid modeling context and, therefore, it can provide a more representative description for the uneven spread of the pandemic.

The descriptive methods which are used in the analysis are graphic methods aiming to display different aspects of distributions of the available data, either in a spatial context (spatial distribution maps, see Fotheringham and Rogerson, 2013), or in a single-variable (boxplots plotting the median, $Q_1$ and $Q_3$ quartiles, and potential outliers and extreme values) or pair-wise (boxplots and scatter-plots plotting ordered pairs of numeric values corresponding to different variables) consideration (Norusis, 2011; Walpole et al., 2012). In terms of statistical inference, the analysis is based on the formulation of error-bars representing confidence-intervals (CIs) that are constructed for estimating (at a 95% confidence level) the difference of the mean values between groups of cases within a variable (Walpole et al., 2012). These error-bars graphically illustrate an independent samples *t*-test of the mean (Norusis, 2011), because when they intersect with the zero-line (horizontal axis), the mean values of the groups cannot be considered as statistically different, whereas when they do not intersect the zero-line can.

Parametric fitting techniques are applied to estimate the parametric curve that best describes the variability of the dataset displayed in a scatter-plot. The available fitting curves that are examined in this part of the analysis are linear (1st-order polynomial, abbreviated Poly1), quadratic (2nd-order polynomial, Poly2), cubic (3rd-order polynomial, Poly3), one-term power-law (Power1), one-term Gaussian (Gauss1), one-term exponential (Exp1), and one-term logarithmic (Log1). All available types of fitting-curves can be generally described by the general multivariate linear regression model (Walpole et al., 2012):

$$\hat{y} = b_1 f_1(x) + b_2 f_2(x) + \ldots + b_n f_n(x) + c = \sum b_i f_i(x) + c \quad (1),$$

where $f(x)$ is either logarithmic $f(x)=(\log(x))^m$, or polynomial $f(x)=x^m$, or exponential $f(x)=(\exp\{x\})^m$, with $m=1,2,3$, or power $f(x)=a^x$. The curve-fitting process estimates the $b_i$ and $a$ (where is applicable) parameters that best fit the observed data and simultaneously minimize the square differences $y_i - \hat{y}_i$ (Walpole et al., 2012), as is shown in the relation:

$$\min\left\{e = \sum_{i=1}^{n}[y_i - \hat{y}_i]^2\right\} = \min\left\{\sum_{i=1}^{n}\left[y_i - \left(\sum b_i f_i(x) + c\right)\right]^2\right\} \quad (2)$$

The parameter estimation is implemented by using the Least-Squares Linear Regression (LSLR) method, which is based on the normality assumption for the differences $e \sim N(0, \sigma_e^2)$ (Norusis, 2011; Walpole et al., 2012).

Finally, the non-parametric kernel density estimation (KDE) method is used to estimate the probability density function of a random variable. The KDE method returns an estimate $\hat{f}(x)$ of the probability density function for the sample data in a vector variable *x*. The estimate is based on a normal kernel function (Botev et al., 2010; Kim and Scott, 2012) and is evaluated at equally-spaced (100 in number) points $x_i$ that cover the data's range. In particular, for a uni-variate, independent, and identically distributed sample $\boldsymbol{x}=(x_1, x_2, \ldots, x_n)$, which is extracted from a distribution with unknown density (at any given point $x$), the kernel density estimator $\hat{f}_h(x)$ describes the shape of the probability-density function $f$, according to the relation (Botev et al., 2010; Kim and Scott, 2012):

$$\hat{f}_h(x) = \frac{1}{n}\sum_{i=1}^{n} K_h(x - x_i) = \frac{1}{nh}\sum_{i=1}^{n} K\left(\frac{x - x_i}{h}\right) \quad (3),$$

where $K$ is the kernel (a non-negative) function and $h > 0$ is a smoothing parameter called bandwidth, which provides a scale (desirably the smallest possible $h$) in the kernel function $K_h(x) = 1/h \cdot K(x/h)$ depending on the bias-variance trade-off dilemma (Geman et al., 1992).

Overall, the multilevel analysis builds on statistical mechanics of the available network, socioeconomic, and geographical variables to conceptualize the worldwide uneven spatio-temporal spread of COVID-19 within the context of the global interconnected economy that is represented by the GTN.

### 3. RESULTS AND DISCUSSION

At the first step of the analysis, we construct the heat-map of Fig.1, which shows the worldwide spatial distribution of the first emergence of COVID-19 per country (variable *DFW* expressing the number of days from Wuhan since the first infection). This heat-map shows some clusters in the world map with distinguishable geographical patterns. The first cluster includes the red-colored cases, expressing cases where the first COVID-19 infection emerged relatively soon after the day the pandemic started in Wuhan (cluster of shortly infected countries). This cluster is distributed mainly throughout the countries neighboring China, along with North America, Australia, and Western Europe. The geographical distribution of this cluster configures a spatial pattern shaping an arc (northern and eastern) consisting of North America - Western Europe - Russian Federation - China-India - Thailand Islands - Australia, which covers the northern and eastern part of the world map.

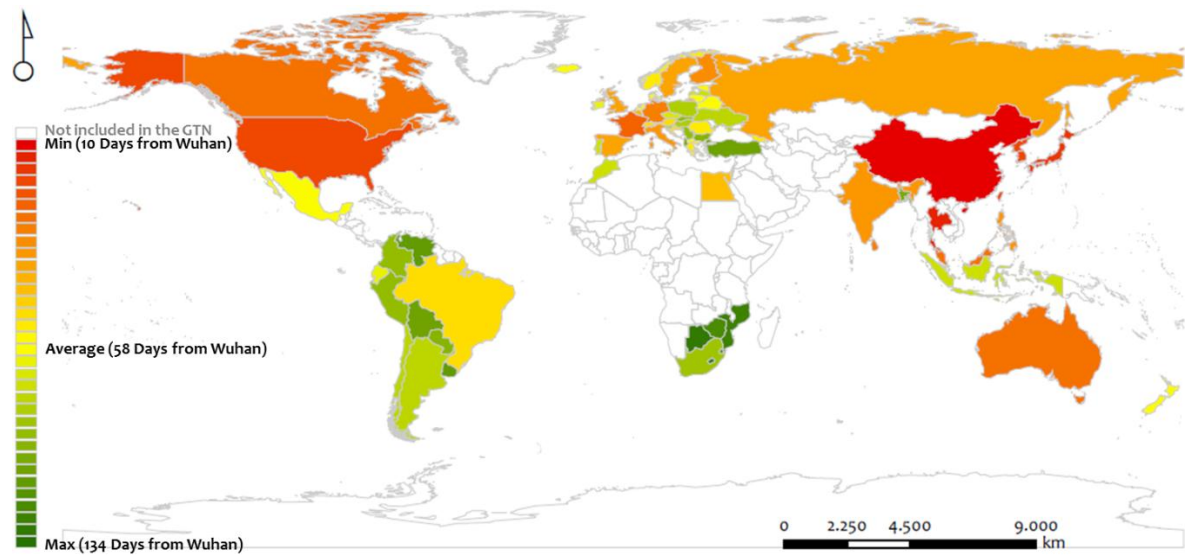

**Fig.1.** Heat-map illustrating the spatial distribution of the temporal spread variable (*DFW*) expressing the number of days from Wuhan (dfW) since the first case emerged in a country (i.e. the days of the first infection per country), for the countries included in the GTN.

The second cluster (Fig.1) includes the green-colored cases, which describe countries where the first COVID-19 infection emerged relatively far from the day the pandemic began in Wuhan. This cluster of late infected countries is mainly distributed along the meridian zone covering the South America, Southern Africa, and Indonesia, and it also includes a sub-cluster of countries in Central Europe and the Western Mediterranean basin, along with Turkey. Finally, the third cluster includes the yellow-colored cases, which describe cases of average emergence of the pandemic from Wuhan (~58 days). This cluster is described by a scattered spatial pattern including European and American countries, which are distributed along a southwestly (in Latin America) and northerly (in Europe) line.

To study the distribution of the first infection per country in terms of the GTN network interconnectedness, in the second part of the analysis, we construct a multilayer scatterplot including boxplot and *ks*-density axes, as is shown in Fig.2. The main panel of the scatterplot (*DFW*, *DEG*≡*k*) shows the correlation between the days since the first infection from Wuhan (dfW) and the node-degree (*k*) of the GTN countries. Boxplots, at the axes, illustrate major aspects of the distributions (median, $Q_1$ and $Q_3$ quartiles, potential outliers, and extreme values) of the corresponding variables (*DFW* and *k*), where, especially on the horizontal axis (measuring days from Wuhan), boxplots are divided into continent groups.

In Fig.2, according to the *ks*-density plot and the pattern of the scatterplot shown in the main panel, we can observe two stages in the COVID-19 temporal spread throughout the GTN. These stages are configured by the distinguishable bell-shaped areas shown in the *ks*-density curve, which are defined by the cutting point of the 44$^{th}$ day from Wuhan (*t*=44dfW). The detection of these stages is a result of the network configuration that applied a filter to the world countries keeping only those 75 belonging to the GTN. In particular, the first stage includes nodes that were infected before the 44$^{th}$ day from Wuhan (≤44dfW) and is mainly described by the outbreak in Asia and North America (as is evident by the country boxplots). The second one includes nodes that were infected after the 44$^{th}$ day from Wuhan (>44dfW) and is described by the outbreak in Europe, South America, and Africa, while the outbreak in Oceania is spread along both stages, but it is slightly positively asymmetric, having its median-value placed at the first stage. Although the pandemic emerged in Europe mainly in the second stage, the cases of the USA (*k*=26), UK (*k*=25), Germany (*k*=22), France (*k*=20), and the Russian Federation (*k*=17) were faced with COVID-19 in the first stage. All these European countries are hubs (i.e. nodes of high degree) in the GTN and they all belong to the $Q_4$ quartile (*t*≤44dfW, $k_i$>15), as is shown in Fig.2.

On the other hand, the late infected nodes mainly concern African countries belonging to the $Q_2$ quartile (*t*>44dfW, $k_i$≤15) and, in terms of the GTN connectivity, they are spokes, namely nodes of one connection nodes, with degree *k*=1. As is evident from the *ks*-density distribution, the majority of nodes (>85%) were infected by the pandemic between the 20$^{th}$ and the 70$^{th}$dfW, where the interquartile range (50% of data) is defined at the period 30-64$^{th}$dfW. The parametric fitting-curve that is applied to the average degree (<*k*>) data shows that the correlation <*k*>=*f*(*t*) is described by a decaying exponential pattern with mathematical expression <*k*>=*f*(*t*)=10.82·exp(-0.018*t*), which implies that the hubs in the GTN are early infected by the pandemic (which is also verified by the fact they belong to the first stage), while lower degree nodes were late infected, on average. In general, this fitting-curve along with the multilayer scatterplot shows that the relation of interconnectedness in the GTN and the day of COVID-19 infection are not likely to be a result of randomness, implying that network interconnectedness is related to the temporal spread of the pandemic within a causative context.

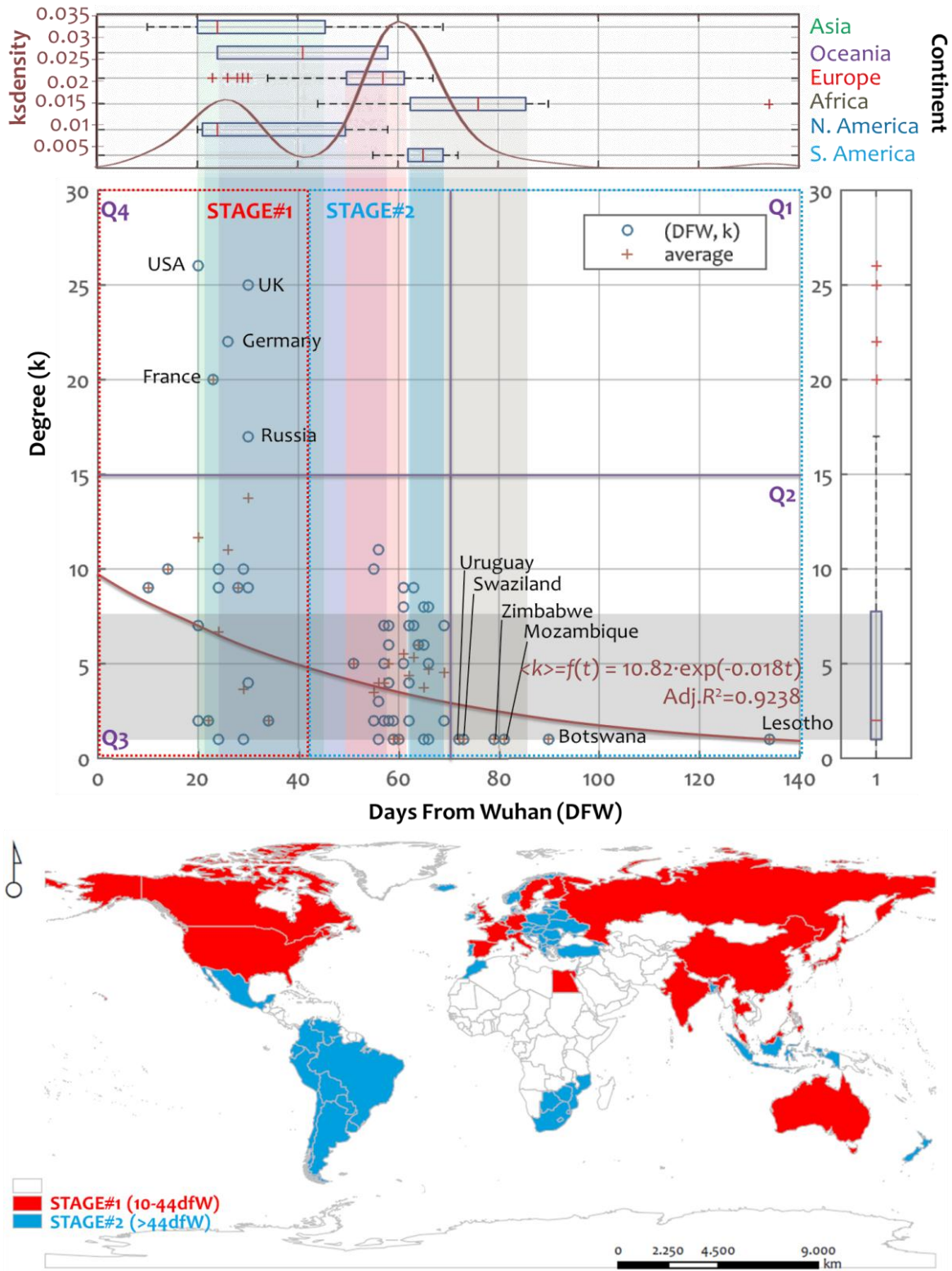

**Fig.2.** Multilayer scatterplot (*DFW,k*) showing the correlation between the days since the first infection from Wuhan (*DFW*) and the node-degree (*k*) of the GTN. Boxplots in each axis illustrate the distribution of each variable, where *DFW* is further separated into continent groups. Shaded zones within the scatterplot express interquartile ranges of each boxplot. Quadrants $Q_1$, $Q_2$, $Q_3$, and $Q_4$ in the scatterplot area are defined by mean-interval values. The fitting curve *f(x)* is applied to average degree values (*<k>*), which are expressed by cross "+" symbols. In the bottom map, the spatial distribution of the two stages defined by the *ks*-density curve is shown.

In geographical terms, the map in Fig.2 illustrates the spatial distribution of the two stages of COVID-19's temporal spread in the GTN. As can be observed, the first stage of the pandemic's temporal spread mainly covers the northern hemisphere, whereas in the second stage it is the southern hemisphere, with notable exceptions being the cases of Central Europe and Australia, respectively. As is evident from the previous analysis, the spatial patterns of the two-stage worldwide temporal spread of COVID-19 in the GTN are more a matter of network interconnectivity (node degree) than of spatial proximity.

To study more in-depth the effect of proximity in the temporal spread of the pandemic throughout the GTN structure, we construct the boxplots of Fig.3, which illustrate how the variables of days since the first infection from Wuhan (Fig.3a) and spatial (geographical) distance (Fig.3b,c) are distributed along groups configured by the node-eccentricity of the GTN. To provide a reference to the case of China, due to its importance in the spread of the pandemic, we center the node-eccentricity to China by subtracting all scores of node-eccentricity by China's eccentricity, which equals 3 steps. Therefore, we compute a new variable named "eccentricity from China" (*ECCFC*), which equals $ECCFC(i)=ECC(i)–3$, where $i$ expresses a node in the GTN.

In terms of descriptive statistics, the boxplots of Fig.3 illustrate the correlations of the pairs of variables (*DFW*, *ECCFC*), (*DSTFC*, *ECCFC*), and (abs(*DSTFC*), *ECCFC*). As can be observed in Fig.3a, the curve fitting applied to the boxplot medians of the eccentricity groups shapes a cubic pattern, which describes the median-data variability under a high level of determination (adj.$R^2$=0.9512). This "U"-shaped pattern yields a global minimum at the value of $ECCFC(i)=0$, which implies that, on average, the countries where the pandemic first emerged are those with the same node-eccentricity as China, in the GTN. In all other cases (nodes), the days since the first infection are almost symmetrically distributed from both sides of the group defined by China's eccentricity. Although it can be claimed on average (and more accurately, to the extent that the median-values are representative of the cases included in a boxplot), this observation implies that the center of the spread of the pandemic in the GTN was not only China, but the core of countries having the same score of eccentricity as China, namely all these nodes (countries) that are as central in this network as China is. As can be observed in the map of Fig.3d, these countries belonging to the eccentricity-core of China are not described by geographical proximity with China, but in terms of network connectivity; they are as central in the GTN as China is.

Moreover, Fig.3b and 3c show that spatial proximity does not seem to be particularly related to the temporal spread of the pandemic (*DFW*), because curve-fitting does not yield any pattern with considerable determination ($R^2$>0.5). Intuitively, the median-values arrangement in the boxplots in Fig.3b and 3c approximates an almost linear pattern in parallel to the horizontal axis, which might imply that the temporal spread of COVID-19 is indifferent to geographical proximity. The overall approach applied in this part of the analysis highlights the importance of network centrality (to the extent that it is described by the metric of eccentricity, which measures central positioning in the network) in the worldwide temporal spread of the pandemic.

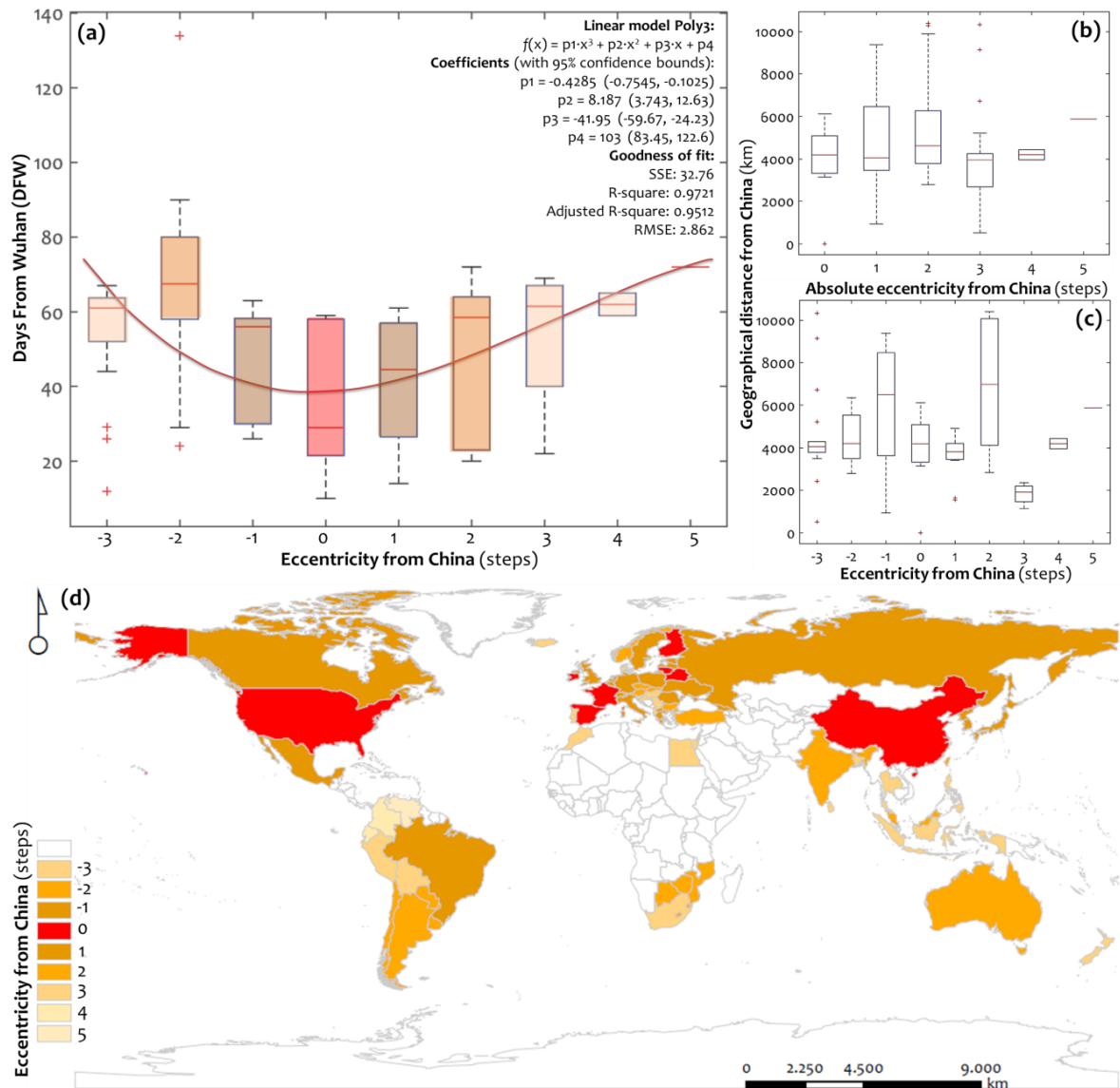

**Fig.3.** Boxplots expressing (a) the days since the first infection from Wuhan (DFW) for each class of the GTN's eccentricity centered at China (abbreviated: eccentricity from China), the eccentricity of which is 3 steps (the fitting curve with the best determination that is applied to the averages is also shown), and the geographical distance from Wuhan (DFW) for each class of (b) absolute and (c) non-absolute eccentricity from China. The fitting curve with the best determination that is applied to the averages is also shown. In the bottom map (d), the spatial distribution of the eccentricity from China is shown.

In the final step of the analysis, we examine which variables included in the 3D-conceptual model of Table 1 can be considered as significant determinants for the worldwide spatio-temporal spread of COVID-19, in the GTN. To do so, we apply a series of *t*-tests to compare the means between the groups defined by the two stages of the temporal spread of the pandemic, as is shown in Fig.1. For better supervision of the results, the *t*-tests are visualized by the error-bars shown in Fig.4, where each variable is standardized to the interval [0,1] so that the results of the *t*-tests are comparable. In cases where the error-bars intersect with the horizontal axis (zero-line), the mean values of the groups can be considered as statistically equal, under a 95% certainty, whereas when they do not intersect the zero-line can be considered as statistically different (suggesting that one group has better performance than the other).

As can be observed in Fig.4, in terms of network interconnectedness (1D conceptual component), the GTN nodes (countries) belonging to the first stage of temporal spread are cases with a higher degree (variable *DEG*, expressing the number of connections of a node in the GTN), outgoing degree (variable *OUT.DEG*, expressing the number of outgoing connections of a node in the GTN), absolute eccentricity from China (variable *EECFC(ABS)*), expressing the network binary distance from China), strength (variable *STR*, expressing the sum of incoming and outgoing tourists a GTN-node annually mobilizes), incoming strength (variable *IN.STR*, expressing the sum of incoming tourists a GTN-node annually receives) and outgoing strength (variable *OUT.STR*, expressing the sum of outgoing tourists a GTN-node annually sends to other destinations). The *t*-tests applied to the variables of this conceptual component indicate that network interconnectivity and central structure are significant determinants in the early global temporal spread of COVID-19. In terms of economic openness (2D conceptual component), the GTN countries belonging to the first stage of temporal spread are cases with a higher globalization index (variable *GI*), GDP and GDP per capita (variable *GDP.pc*), and total factor productivity per capita (variable *TFP.pc*) than the GTN nodes that are included in the second stage of temporal spread. The *t*-tests applied to the variables of this conceptual component show that the countries with higher economic openness (i.e. those that are more integrated into the globalized economic structure) were subjected earlier to the infection of the pandemic than the countries of lower economic openness.

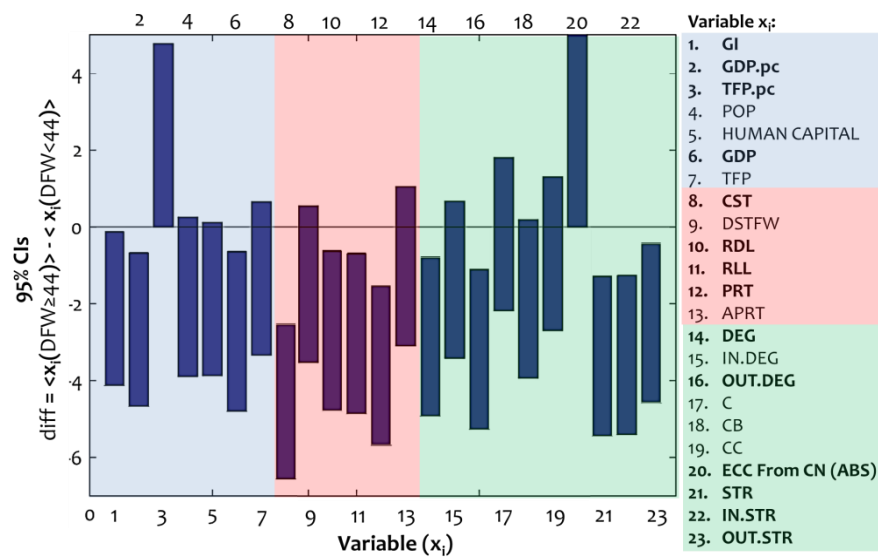

**Fig.4.** Error bars of 95% confidence intervals (CIs) for the average differences $\langle \text{diff}(x_i) \rangle$, which are computed on standardized variables (i.e. ranging from 0 to 1) between the groups of cases $x(t<42\text{DFW})$ and $x(t \geq 42\text{DFW})$ defined by the cutting value $t=42\text{DFW}$ (the number of days since the first infection in Wuhan), for each of the available network and economic variables ($x_1, x_2, \ldots, x_{19}$). Labels shown in bold-font have significant differences.

Finally, in terms of spatial impedance (3D conceptual component), the GTN countries belonging to the first stage of temporal spread are cases with more coastal geomorphology (variable *CST*), they have larger road (variable *RDL*) and rail lengths (variable *RLL*), and a greater number of ports (variable *PRT*) than those included in the second stage of temporal spread. An interesting insignificant result that can be observed in this conceptual group is the variable APRT describing the number of airports included in the GTN countries. This observation, in conjunction with the significant *t*-tests observed for the degree (*DEG*) and strength (*STR*) variables (included in the network interconnectedness group), implies that the airport network becomes a significant determinant in terms of connectivity rather than in terms of infrastructure capacity. Regardless of the participation of the APRT variable, the *t*-tests applied to the variables of the 3d conceptual component illustrate how land and maritime transport capacity (which are major aspects of transport integration) significantly affected the early temporal spread of the pandemic worldwide. Overall, the *t*-test analysis shows that network interconnectedness, economic openness, and transport integration are key determinants in the early global temporal spread of the pandemic.

## 4. CONCLUSIONS

This paper has developed a multilevel methodological framework for understanding the uneven spatio-temporal spread of COVID-19 in the context of the global interconnected economy. The framework is built on a three-dimensional conceptual model, incorporating one dimension for approximating the interconnectedness in worldwide human mobility, a second one for the global economic openness, and a third one for the spatial impedance to transportation. The analysis was applied to a major temporal variable expressing the day from Wuhan since the first infection and to another twenty-three variables that were grouped into the categories of the 3D conceptual approach. Firstly, the descriptive analysis revealed three clusters in the world map with distinguishable geographical patterns of the pandemic temporal spread. The first one of early infected countries configured a geographical arc distributed throughout the countries neighboring China, North America, Australia, and Western Europe. The second cluster of late infected cases was mainly distributed along the meridian zone of South America, Southern Africa, and Indonesia, also including a sub-cluster with countries of Central Europe, the Western Mediterranean basin, and Turkey, while the third one configured a scattered spatial pattern throughout the globe. The parametric and non-parametric statistical analysis led to further specialization of these findings and revealed two main stages in COVID-19's temporal spread throughout the GTN. The first one included nodes that were infected by the pandemic before the $44^{th}$ day from Wuhan ($\leq$44dfW) and described the outbreak in Asia and North America. The second one included nodes that were infected after the $44^{th}$ day from Wuhan (>44dfW) and described the outbreak in Europe, South America, and Africa, while the outbreak in Oceania was spread along both stages. In geographical terms, the first stage of the pandemic's temporal spread appeared to be more a matter mainly of the northern hemisphere, whereas the second stage involved the southern hemisphere. Next, the analysis that was applied to the average degree and temporal spread of the pandemic showed a decaying exponential pattern, indicating that hubs in the GTN were early infected, while lower degree nodes were infected late by the pandemic. This pattern showed that network interconnectedness is related to the temporal spread of COVID-19 within a causative context. Further, the descriptive and parametric statistical analysis revealed a "U"-shaped pattern describing the correlation between network eccentricity and the temporal spread of the pandemic and implied that the center of the pandemic's temporal spread in the GTN included China and the core of countries that are as central in the network as China is. This analysis also revealed that spatial proximity was not a major determinant of the temporal spread of the pandemic. Finally, the statistical inference analysis that was applied to the total set of available variables indicated that network interconnectivity and central structure are significant determinants of the early temporal spread of COVID-19, that countries with higher economic openness were earlier submitted to the infection of the pandemic, and that land and maritime transport integration significantly affected the early temporal spread of the pandemic worldwide. Overall, this paper has revealed two major stages in the temporal spread of the pandemic within the context of interconnected worldwide mobility networks. It highlighted the importance of network centrality in the worldwide temporal spread of the pandemic, it showed that network interconnectedness, economic openness, and transport integration are key determinants in the early global spread of the pandemic, and it revealed that the spatio-temporal patterns of the worldwide spread of COVID-19 were more a matter of network interconnectivity than of spatial proximity.